\documentclass[vecphys]{svmult}

\usepackage{makeidx}
\usepackage{graphicx}
\usepackage{multicol}
\usepackage[bottom]{footmisc}

\usepackage{amsmath}
\usepackage{amsfonts}
\usepackage{graphics}
\usepackage{subfigure}
\usepackage{epsfig}
\usepackage{rotate}
\usepackage{ifpdf}
\usepackage{graphicx,amssymb}

\makeindex
\begin{document}

\title*{Boxy/peanut bulges : formation, evolution and properties}

\author{E.~Athanassoula\inst{1}, I. Martinez-Valpuesta\inst{1,2}}

\institute{Laboratoire d'Astrophysique de Marseille (LAM), 
Observatoire Astronomique \\de
Marseille-Provence (OAMP), 2 Place Le Verrier, 13248 Marseille,
C\'edex \\04, France.\\ \texttt{lia@oamp.fr} \and
IAC, C/Via Lactea s/n, 38200, La Laguna, Tenerife, Spain. \\ \texttt{imv@iac.es}}

\maketitle
\begin{abstract}
We discuss the formation and evolution of boxy/peanut bulges (B/Ps)
and present new simulations results. Orbital structure studies show that B/Ps
are {\it parts} of bars seen edge-on, they have their origin in  
vertical instabilities of the disc material and they are somewhat shorter
in extent than bars. When the bar forms it is
vertically thin, but after a time of the order of a Gyr it experiences
a vertical 
instability and buckles. At that time the strength of the bar decreases, 
its inner part becomes thicker, so that, seen edge-on, it acquires 
a peanut or boxy shape. A second buckling
episode is seen in simulations with strong bars, accompanied by a further
thickening of the B/P and a weakening of the bar. Quantitatively,
this evolution depends considerably on the properties of the halo and
particularly on the extent of its core. This influences the amount of
angular momentum exchanged within the galaxy, emitted by near-resonant
material in the bar region and absorbed by near-resonant material 
in the halo and in the outer disc. Haloes with small cores generally
harbour stronger bars and B/Ps and they often witness double buckling.
\end{abstract}

\section{Introduction}

Disc galaxies viewed edge-on often show in their central parts a
characteristic thickening, which has the shape of a box, of a peanut,
or of an `X' (e.g. \cite{LuttickeDP00a, LuttickeDP00b}). Since these
structures swell out of the disc plane, they are called 
bulges, and more specifically, boxy bulges, or peanut bulges, or `X'
shaped bulges, or, for short, B/Ps or B/P bulges. Yet their
properties, as well as their formation and 
evolution are very different from those of classical bulges
\cite{A05bulges}. In fact, evidence from many studies has shown that
they are just {\it parts} of bars seen edge-on. Here, we will briefly review
relevant results about their orbital structure (Section
\ref{sec:orbits}) and describe their 
formation and evolution as witnessed in $N$-body simulations (Section
\ref{sec:formation}). Based on the inner halo structure, we distinguish
different types of models and discuss the role of the
halo on the growth of the bar and of the B/P.

\section{Orbital structure}
\label{sec:orbits}

In order to understand the dynamics of any structure, it is essential
to study first the periodic orbits that form its backbone. For
two-dimensional bars, the backbone is the well studied family of x$_1$
orbits \cite{ContopPapayan, ABMP}. These 
are elongated along the bar and have an axial ratio that varies both
as a function of their Jacobi constant and of the properties of the
bar potential used \cite{A92a}. When stable, they trap around them other
orbits. The problem is more complex in three dimensions, since a
vertical instability of parts of this family \cite{Binney78} introduces a
number of other families extending vertically well outside the disc,
such as as the x$_1$v$_1$, x$_1$v$_2$, x$_1$v$_3$ etc.
\cite{Pfenniger84, SPA02a, SPA02b, PSA02}. These families, together
with the main 
x$_1$ family from which they bifurcate, are known as the x$_1$ tree. 
\cite{SPA02a}. They are linked to the $n:1$ vertical resonances
and extend well outside the disc equatorial plane. As shown in 
\cite{PSA02}, 
some of them are very good building blocks for the formation of
B/Ps, because their orbits are stable and have the
right shape and extent. Studies of these orbits reproduced many
of the B/P properties and helped explaining crucial aspects of B/P
formation and evolution. For example, an analysis of the orbital
families that constitute B/Ps predicts that they should
be shorter than bars. This is indeed verified both in $N$-body
simulations and in real galaxies (\cite{A05bulges}, 
\cite{LuttickeDP00b}, \cite{AthanassoulaBeaton05}). Furthermore, such
studies predict that stronger bars should correspond to stronger B/Ps,
and this also is verified both in $N$-body simulations 
(\cite{A06Ishigaki} and Athanassoula \& Martinez-Valpuesta, in
preparation) and in real galaxies (\cite{LuttickeDP00b}).  
 
\section{Formation and evolution of boxes and peanuts}
\label{sec:formation}

The formation of boxy/peanut bulges has been witnessed in a large number of
simulations (\cite{CombesSanders81},
\cite{CDFP90}, \cite{RahaSJK91},
\cite{AM02},
\cite{Athanassoula03}, \cite{A05bulges},
\cite{ONeilDubinski03}, 
\cite{DebattistaCMM04}, 
\cite{MartinezVShlosman04}, 
\cite{DebattistaCMMWQ06}, 
\cite{MartinezVSH06}, 
etc). These have many aspects in common, but also many
differences. We will describe and illustrate here two different
characteristic types of evolution, corresponding to two simulation types
which in \cite{AM02} have been labelled MH and MD, respectively.

In both cases the simulation starts with an exponential disc of unit
mass and unit 
scale-length. It is immersed in a live halo and the
halo-to-disc mass ratio is 5. In MH type
models the halo has a small core, smaller or of the order of the disc
scale-length, i.e. it is centrally
concentrated. Thus, in the disc region, the halo contribution to the
circular velocity curve is of the same order as that of the
disc. On the contrary, in MD type models the halo has a very big core,
much larger than the disc scale length, so that in the inner parts the
disc dominates. 

The evolution of the bar in
these two types of models is quite different (\cite{AM02}). In both cases
the bar grows by exchanging angular momentum with the outer disc and
with the halo. Angular momentum is emitted by near-resonant material
in the bar region and is absorbed by near-resonant material in the
outer disc and in the halo (\cite{Athanassoula02}). The halo density at
the locations 
of the resonances is much larger in MH types than in MD types. Thus,
provided the resonances are responsive, there will be more angular
momentum exchanged within the galaxy in MH types than in MD types.
This leads to strong bars in MH types and much weaker ones in MD types
(\cite{AM02, Athanassoula03}).

\begin{figure}[!ht]
%\vspace{2.9in}
\centering
\includegraphics[height=5.2in]{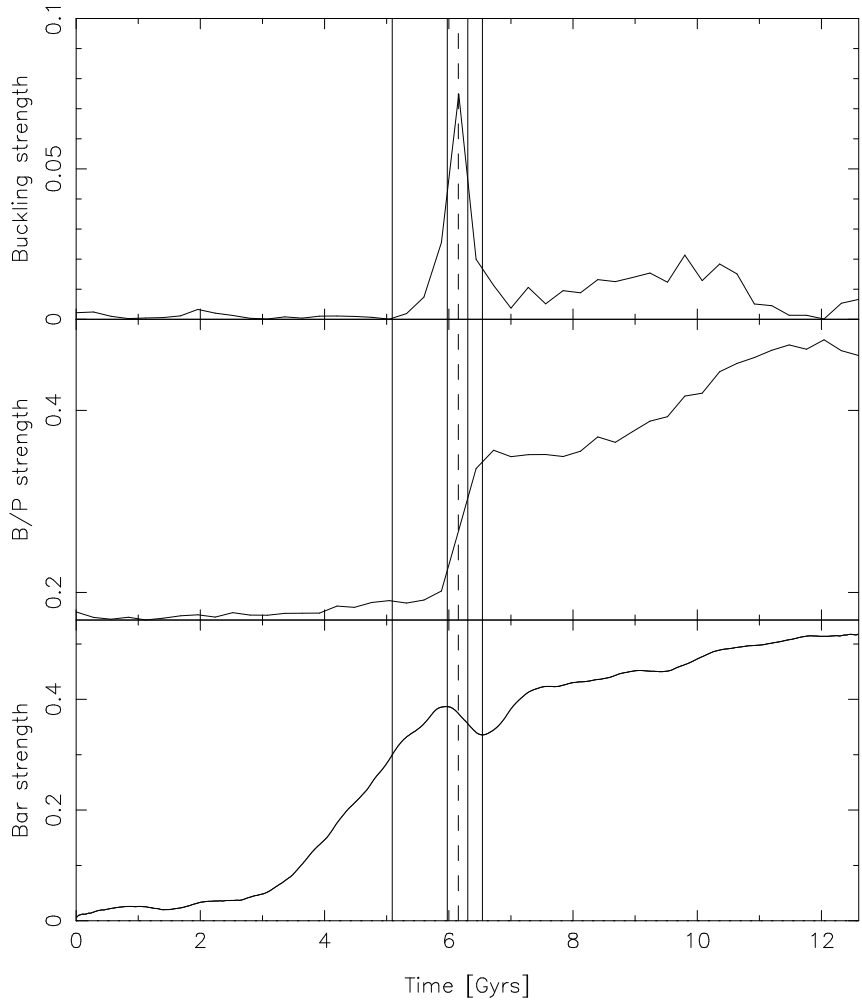}
\caption[]{ Example of the time evolution of three B/P-, or
  bar-related quantities for an MH type model. These are
  the buckling strength (i.e. the vertical asymmetry; upper
  panel), the B/P strength (i.e. its maximum vertical extent; middle 
  panel) and the bar strength (lower panel). The  
  vertical lines mark characteristic times linked to bar
  formation and evolution. From left to right, these are the bar
  formation time, the maximum amplitude time, the bar decay time and
  the bar minimum amplitude time (see text). The vertical dashed line
  marks the time of the buckling.}
\label{fig:tevlMH}
\end{figure}

\begin{figure}[!ht]
%\vspace{0.55in}
\centering
\includegraphics[height=5.2in]{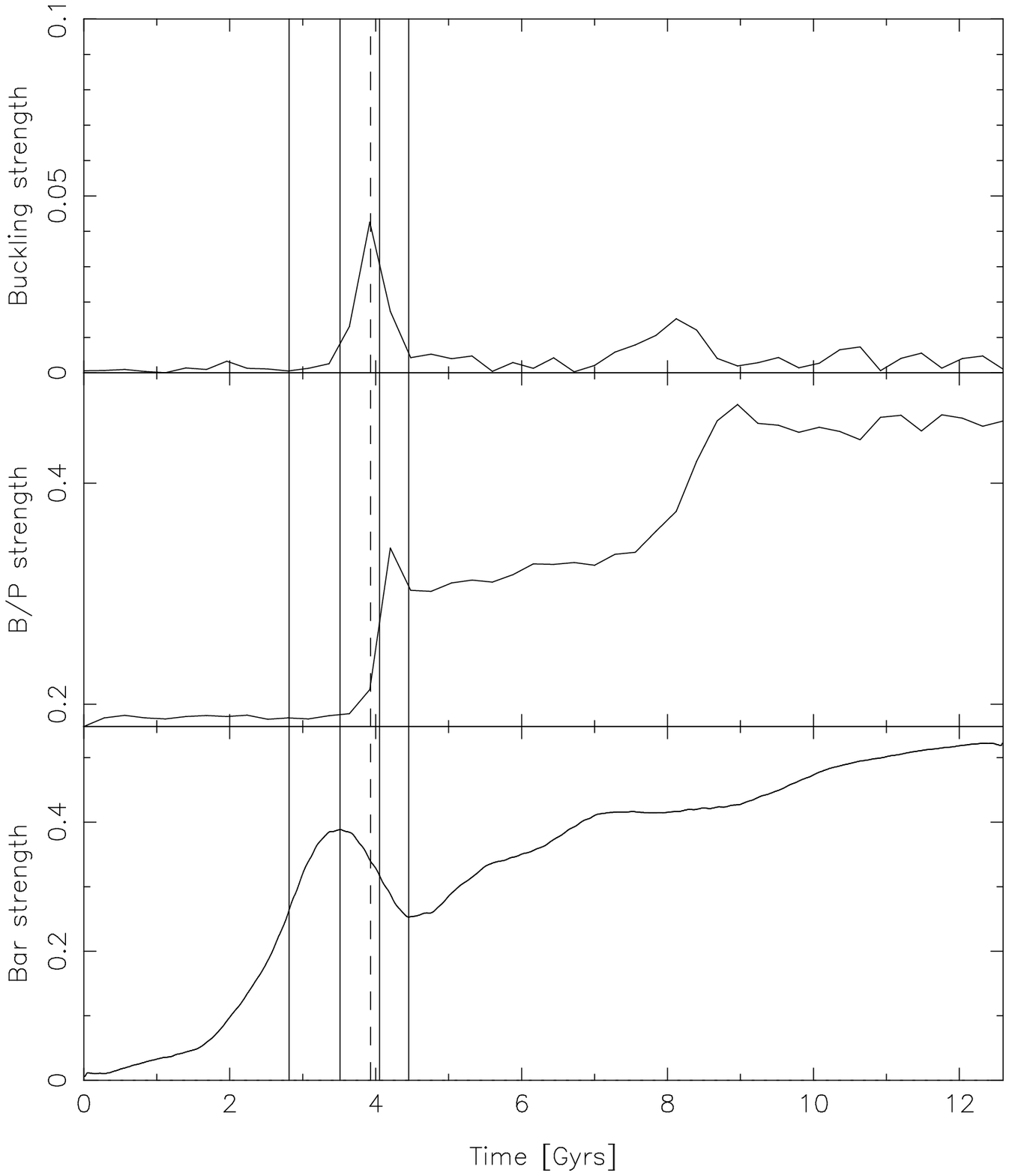}
\caption[]{ As in Fig.~\ref{fig:tevlMH}, but for an MH-type simulation
  with two buckling episodes.
}
\label{fig:tevlMH2}
\end{figure}

\begin{figure}[!ht]
%\vspace{0.55in}
\centering
\includegraphics[height=5.2in]{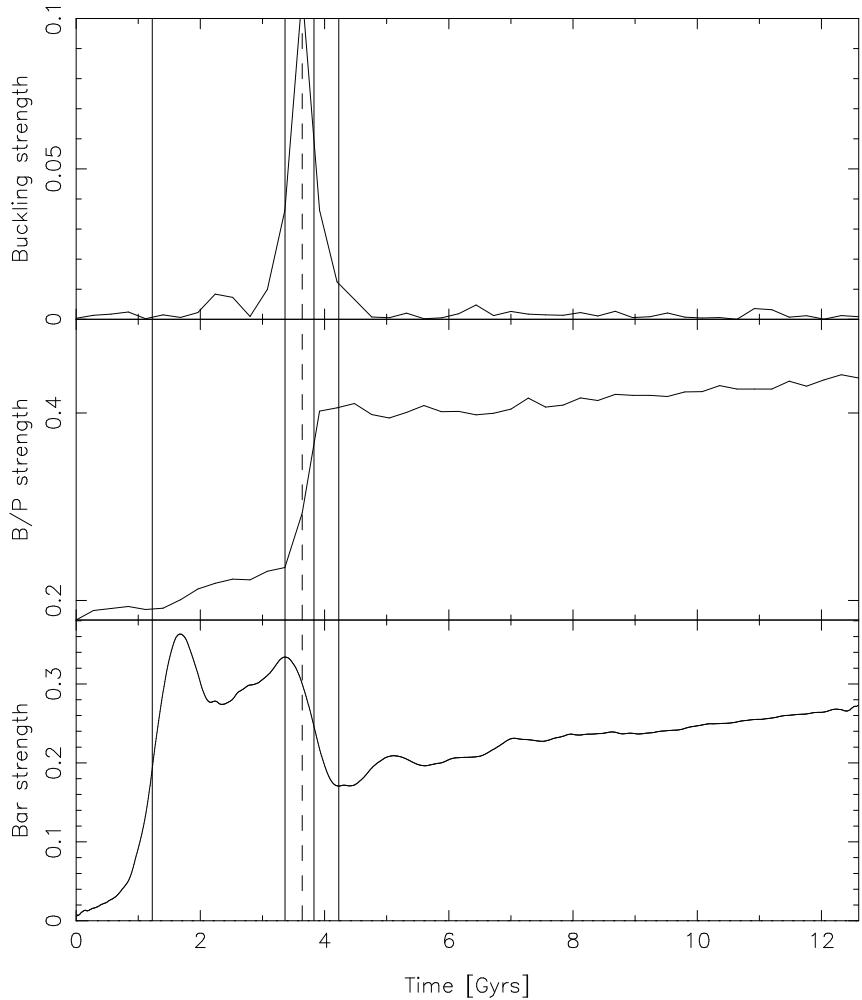}
\caption[]{ As in Fig.~\ref{fig:tevlMH}, but for an MD-type simulation.}
\label{fig:tevlMD}
\end{figure}

The time evolution of characteristic properties of the bar and the B/P
is shown in Figs.~\ref{fig:tevlMH}, \ref{fig:tevlMH2} and
\ref{fig:tevlMD} for two MH and 
one MD type simulations, respectively. From top to bottom, the panels
give the buckling strength, the B/P strength and the bar strength.  
The time is given in Gyrs, using the calibration proposed in
\cite{AM02}. Fig.~\ref{fig:tevlMH} shows that
the initially unbarred disc forms a bar roughly between times 3 and
6 Gyrs (lower panel). We define as bar formation time the time at which
the bar-growth 
is maximum (i.e. when the slope of the bar strength as a function of
time is maximum) and indicate it by the first vertical line in
Fig.~\ref{fig:tevlMH}. The bar strength reaches a maximum at a time
noted by the second vertical line, and then decreases considerably over
$\sim$ 0.5 Gyr. The time at which the bar amplitude decrease is maximum is
given by the third vertical line. Subsequently, the bar strength reaches a
minimum, at a time shown by the fourth vertical line, and then starts
increasing again at a rate much slower than that during bar
formation. 

The upper panel shows the buckling strength, i.e. the vertical 
asymmetry, as a function of time. Before the bar forms the disc is
vertically symmetric, with the first indications of asymmetry occurring
after bar formation. The asymmetry grows very abruptly to a
strong, clear peak and then drops equally abruptly. The time of the
buckling (dashed vertical line) is very clearly defined as the maximum
of this curve. The middle panel shows the strength of the B/P, i.e. its
maximum vertical extent, again as a
function of time. This quantity grows abruptly after the bar
has reached its maximum amplitude and during the time of the buckling.
This abrupt growth is followed by a much slower increase over a longer
period of time. 

Taken together, the three panels of
Fig.~\ref{fig:tevlMH} show that the bar forms 
vertically thin, and only after it has reached a maximum strength does the
buckling phase occur. During the buckling time the bar strength
decreases significantly, while the B/P strength increases. The time
interval during which B/P formation, or buckling occur is rather
short, of the order of a Gyr, and it is followed by a
longer stretch of time during which the bar and B/P evolve much
slower. 

Fig.~\ref{fig:tevlMH2} shows results for another MH simulation. The
first part of the 
evolution is very similar to that of the previous example, except that
the time for bar formation is shorter and the time during which the bar
amplitude decreases is somewhat longer. This example, however, has a very
interesting feature : it has a second, weaker buckling episode
shortly after 8 Gyrs. This occurs very often in simulations developing strong
bars and was discussed first in \cite{A05gdansk} and \cite{MartinezVSH06}. 
It is seen clearly
in all three panels and has characteristics similar to those of the
first buckling. Namely, there is an asymmetry, with a clear maximum of
the buckling strength, a sharp increase of the B/P strength and a
flattening of the bar strength. In many other examples with two
buckling episodes, instead of a
flattening there is a decrease of the bar amplitude. Thus, the main
differences between the two buckling episodes are only that the 
buckling peak is broader and less high and that the drop in the bar
strength is less strong. The time between the two bucklings, in this
example, is about 4 Gyrs.

The evolution of the same quantities for the MD simulation is given in
Fig.~\ref{fig:tevlMD}. Many crucial differences are immediately
clear. The bar forms much faster than for the MH type models, in less
than 2 Gyrs. This is in good agreement with results of previous
simulations \cite{AthanassoulaSellwood86, Athanassoula02}. 
There is a first maximum of the bar amplitude before 2
Gyrs, which, however, is not associated with a maximum of the
asymmetry (buckling), nor with a sharp increase of the B/P
strength. It thus can not be linked to the B/P formation, and we
checked this further by viewing the evolution of the simulation by
eye. This revealed that in the initial stages of the simulation a very
long and thin bar 
forms, due to the strong instability in the disc-dominated inner
region. This bar drives a strong, two-armed spiral which heats the
disc and thus lowers its own amplitude. Furthermore, the bar
is so thin that it must render many of the orbits that support it
chaotic \cite{ABMP}, so that they can not support it
further. Thus, the bar strength should decrease spectacularly and this
is indeed witnessed in Fig.~\ref{fig:tevlMD}. Subsequently, the bar
amplitude increases with time until the formation of the B/P, which, as
in the previous examples, is clearly seen as a maximum of the buckling
strength, a sharp increase of the B/P strength and a sharp decrease of
the bar strength. The time between the bar growth and its decay is
much longer than in the previous examples. In this specific case it is
about 2.5 Gyrs, but in other cases it can be considerably longer. 

It is thus clear that the halo properties, and in particular the
size of its core, influence strongly the time evolution of the bar and
of the B/P.

%===============================================
\section*{Acknowledgements}
We thank A. Bosma for interesting and fruitful
discussions. This work has been partially supported by grant 
 ANR-06-BLAN-0172 and by the Gr\"uber foundation.

\end{document}